\documentclass[epj,twocolumn]{webofc}
\usepackage[varg]{txfonts}   
\usepackage{amsmath}
\usepackage{amssymb}
\usepackage{slashed}
\usepackage{physics}
\usepackage{xcolor}

\newcommand{\valencia}{Instituto de F\'{\i}sica Corpuscular, Universitat de Val\`{e}ncia -- Consejo Superior de Investigaciones Cient\'{\i}ficas, \\Parc Cient\'{\i}fic, \mbox{E-46980} Paterna, Valencia, Spain.}

\woctitle{Flavour changing and conserving processes}
\begin{document}
\title{NNLO QED contribution to the $\mu e\to \mu e$ elastic scattering}
%
%


\author{Jonathan Ronca\inst{1}\fnsep\thanks{
           Based on a collaboration with S. di Vita, S. Laporta, P. Mastrolia, M. Passera, T. Peraro, A. Primo, U. Schubert and W.J. Torres Bobadilla. } \thanks{\email{Ronca.Jonathan@ific.uv.es}}
}

\institute{\valencia}

\abstract{
We present the current status of the Next-to-Next-to-Leading Order QED contribution to the $\mu e$-scattering. Particular focus is given to the techniques involved to tackle the virtual amplitude and their automatic implementation. Renormalization of the amplitude will be also discuss in details.
}
\maketitle
\section{Introduction}

The current measurements of the muon anomalous magnetic moment $a_\mu=(g_\mu-2)/2$ indicate a discrepancy of $\sim 3.5\sigma$ from the Standard Model prediction \citep{blum2013muon}. This fact might be a hint of new physics. New upcoming experiments at Fermilab and J-PARC will improve significantly the precision of $a_\mu^{exp}$, hence higher accuracy for $a_\mu^{th}$ from the theoretical side is required.

A very clean way to achieve the aimed precision involves the determination of the leading hadronic contribution $\Delta\alpha^{\text{had}}(q^2)$ to the electromagnetic coupling in the space-like region \citep{Carloni_Calame_2015,Abbiendi_2017}. 
The \emph{MUonE experiment}, recently proposed at the CERN, is meant to obtain an estimation of $\Delta\alpha^{\text{had}}(q^2)$ through a measurement of the differential cross section of the $\mu e$-elastic scattering. In order to be competitive with the time-like datas, a precision of 10ppm is required, both on statistical and systematic errors. 
 
QED contributions to $\mu e$-elastic scattering constitute an irreducible background for the measurement, and they represent a relevant source of systematic errors; results for Leading Order (LO) and Next-to-Leading Order (NLO) corrections are already available \citep{Alacevich:2018vez}, as well as the Next-to-Next-to-Leading Order (NNLO) hadronic ones \citep{Fael:2018dmz, Fael:2019nsf}. However, Next-to-Next-to-Leading Order (NNLO) corrections to $\mu e$-elastic scattering are mandatory to achieve the 10ppm precision.
 

In this work, we summarize the current status of the double-virtual contributions  to the NNLO QED $\mu e$-elastic scattering. The review is organized as follows: Section 2 recaps the external kinematic definitions we use for characterizing the $\mu e$-elastic scattering; Section 3 states the basics of the calculation process, stressing the importance of the \emph{Feynman integrals}. The next sessions will present the heart of this calculation: the \emph{decomposition} of the amplitude and the evaluation of the \emph{master integrals}.


The decomposition strategies are presented in Section 4. We discuss the main features of the Adaptive Integra\emph{nd} Decomposition \citep{Mastrolia_2016,mastrolia2016adaptive} and its implementation \textsc{Aida} \citep{primophdth,torresphdth}, and the Integra\emph{l} Reduction algorithms involving the \emph{Integration-by-parts identities} \cite{Chetyrkin:1981qh,Laporta_2000,manteuffel2012reduze,Maierhoefer:2017hyi}. These method are exploit to obtain a reduction of the full amplitude in terms of a minimal set of master integrals. Section 5 is focused on the evaluation approaches, presenting the analytical and the numerical ways. Analytical strategy is based on solving \emph{systems of differential equations} \citep{Barucchi_1973,Kotikov:1990kg,Remiddi:1997ny,Gehrmann_2000} which master integrals obey. The main results of the analytical method for the $\mu e$-elastic scattering are collected into \citep{Mastrolia_2017,Di_Vita_2018}, and cross-checked numerically.
Alternatively, a numerical method which can be userful in these calculation is the so-called \emph{sector decomposition} \citep{HEINRICH_2008}, which will be briefly introduced. 

Section 6 is devoted to present the complete \textsc{Mathematica} implementation which embeds each step of the calculation, from the generation of the amplitude to evaluation of the master integrals. A flowchart of the global algorithm is presented in Fig. \ref{fig}.

In Section 7 we define the renormalization procedure which will be adopted to obtain a UV finite $\mu e$-elastic Scattering Amplitude. Both $\overline{\text{MS}}$ and \emph{on-shell} schemes are employed to renormalize respectively the QED coupling constant and the muon fermion field and mass.

Lastly, we present the main result achieved for the $\mu e$-elastic scattering  and we point out the next steps needed to complete the full NNLO contribution.

\section{$\mu e$-elastic Scattering process}
The process under investigation is the elastic scattering
\begin{equation}
\mu^+(p_1) e^-(p_2) \to e^-(-p_3)\mu^+(-p_4),
\end{equation}
where $p_i$ are the on-shell momenta: $p_1^2=p_4^2=m_\mu^2$ and $p_2^2=p_3^2=m_e^2$.
Let $p_4$ be the dependent momentum. From the general Mandelstam variables definition, \mbox{$s_{ij}=(p_i+p_j)^2$}, define 
\begin{equation}
\begin{gathered}
s=s_{12},\quad t=s_{23},\quad u=s_{13},\\
s+t+u=2m_e^2+2m_\mu^2.
\end{gathered}
\end{equation}
The physical phase-space region is constrained by the following conditions
\begin{equation}
\begin{gathered}
s\ge (m_e+m_\mu)^2,\\
-\frac{\lambda(s,m_\mu^2,m_e^2)}{s}<t<0,\\
\begin{aligned}
\lambda(s,m_\mu^2,m_e^2)~=~&
s^2+(m_\mu^2)^2+(m_e^2)^2-2sm_\mu^2\\
&-2sm_e^2-2m_e^2m_\mu^2.
\end{aligned}\end{gathered}
\end{equation}

The ratio between electron and muon mass is of the order $10^{-5}$. This enforce the validity of the massless electron approximation, which yields many advantages to the NNLO calculation. Therefore, from now on we set $m_e=0$ and $m_\mu=m$.

\section{Double-virtual contributions}
Within the perturbative QFT approach, the QED cross section $\sigma$ of the $e\mu$-scattering is a series expansion in terms of the coupling constant $\alpha$,
\begin{equation}\label{sig1}
\sigma=\sigma_{LO}+\sigma_{NLO}+\sigma_{NNLO}+\cdots+\mathcal{O}(\alpha^{2+n}).
\end{equation}
For $2\to 2$ scattering process, $\sigma$ is also related to the Feynman amplitude $\mathcal{M}$ through the well-known relation
\begin{equation}\label{sig2}
\sigma=C_{\text{flux}}\int |\mathcal{M}|^2 d\text{PS}_{2},
\end{equation}
where $C_{\text{flux}}$ is the flux factor and $d\text{PS}_{2}$ is the infinitesimal phase-space element. Matching Eq. (\ref{sig1}) and (\ref{sig2}), the order-by-order expression of $\sigma$ is manifest,
\begin{equation}\label{completecalc}
\begin{aligned}
\sigma_{LO}\sim &~ |\mathcal{M}_{0}|^2, \\
\sigma_{NLO}\sim &~ 2\text{Re}[\mathcal{M}_{0}^*\mathcal{M}_{1l}]+|\mathcal{M}_{\text{r}}|^2,\\
\sigma_{NNLO}\sim
&~ 2\text{Re}[\mathcal{M}_{0}^*\mathcal{M}_{2l}]+2\text{Re}[\mathcal{M}_{1l}^*\mathcal{M}_{1l}]\\
&+2\text{Re}[\mathcal{M}_{r}^*\mathcal{M}_{1l,r}]+|\mathcal{M}_{2r}|^2, 
\end{aligned}
\end{equation}
where $\mathcal{M}_{0}$ is the Born amplitude, $\mathcal{M}_{1l}$ and $\mathcal{M}_{2l}$ are respectively the one and two-loop Feynman amplitude, $\mathcal{M}_{r}$ and $\mathcal{M}_{2r}$ are the real and the double-real emission amplitudes, with respectively one and two more particles emitted in the final state, and $\mathcal{M}_{1l,r}$ the real-virtual amplitudes, one-loop amplitude with one more particle emitted in the final start.

\begin{figure*}
\centering
\begin{center}
\includegraphics[width=\textwidth]{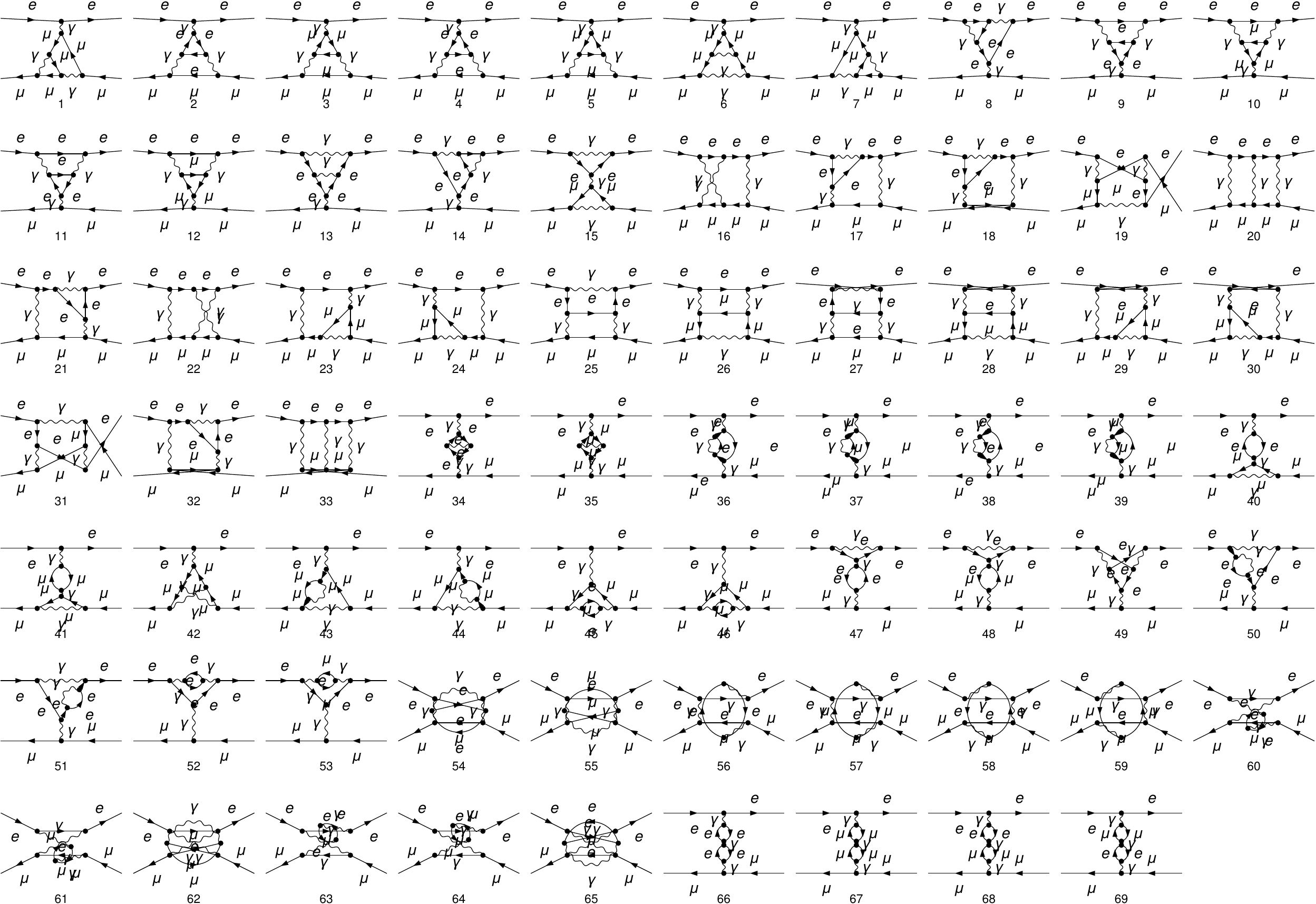}
\caption{Two-loop diagrams contributing the QED NNLO amplitude for the $\mu e$-scattering}\label{diags}
\end{center}
\end{figure*}

The double-virtual contribution  $\mathcal{M}_{2l}$ to $\sigma_{NNLO}$ (Fig. \ref{diags}) is a combination of dimensionally regularized \emph{Feynman integrals}
\begin{equation}\label{amplfull}
2\text{Re}[\mathcal{M}_{0}^*\mathcal{M}_{2l}]=\sum_{\bar{i}}A_{\bar{i}}I_{\overline{i}}(\mathbf{s}),
\end{equation}
where the coefficients $A_{\bar{i}}$ depend on the kinematic variables $\mathbf{s}=(s,t,m^2)$.
The Feynman integrals\footnote{$\int_{\mathbf{k}}=\int_\mathbb{M}\frac{d^dk_1}{(i\pi^{d/2})}\frac{d^dk_2}{(i\pi^{d/2})}$, and $\mathbf{k}=(k_1,k_2)$} $I_{\overline{i}}(\mathbf{s})$ have the general form
\begin{equation}
I_{\overline{i}}(\mathbf{s})=\int_{\mathbf{k}}\frac{\mathcal{N}_{\bar{i}}(\mathbf{k},\mathbf{s})}{D^{i_1}_1\cdots D^{i_n}_n},\quad \bar{i}\in\mathbb{N}^n,
\end{equation}
and depend on the kinematic invariants $\mathbf{s}$ and the dimension $d$. The numerator $\mathcal{N}_{\bar{i}}$ is a general polynomial of the scalar products between loop and external momenta. The denominators $D_j$ are inverse scalar propagators, whose form is $D_j=q_j^2(\mathbf{k},\mathbf{p})-m^2_j$. The multi-index $\bar{i}=(i_1,\cdots,i_n)$ contains the exponents of the denominators.

Direct integration of Feynman integrals objects is usually not allowed. A viable way to calculate them efficiently involves decomposition methods: Feynman integrals can be expressed into combinations of "simpler" integrals. Such integrals can be evaluated either analytically of numerically. In the next session, the algorithm employed in the calculation of the amplitude (\ref{amplfull}) is presented.

\section{Decomposing the Amplitude}\label{dec}

There exist several decomposition methods, which can be apply either to integra\emph{nds} \citep{Ossola_2007,Mastrolia_2012,Badger_2012} or integra\emph{ls} \citep{Chetyrkin:1981qh,TKACHOV198165,Passarino:1978jh,Smirnov_2006}. An optimized algorithm to achieve a complete decomposition of a Feynman integrals will involve both methods.

\subsection{Adaptive Integrand Decomposition}

As extensively discussed in \cite{Mastrolia_2016,mastrolia2016adaptive}, the \emph{Adaptive Integrand Decomposition}  (AID) is an algorithm that works at integrand level. It exploits the (rational and) polynomial structure of the integrand in order to achieve an iterative decomposition of the numerator.

The basic ingredient is the splitting of the $d$-dimensional space into \emph{parallel} and \emph{transverse} component \cite{vanNeerven:1983vr} w.r.t. the space spanned by the independent external momenta:
\begin{equation}\label{aid1}
d=d_\parallel+d_\perp \implies k_i^\alpha=k_{\parallel i}^\alpha+\lambda^\alpha_i.
\end{equation}
It can be shown that Eq. (\ref{aid1}) makes the polynomial structure of the integrand manifest, and liable to perform polynomial division without passing by the \emph{Gr\"obner basis} \citep{Mastrolia_20122}. Therefore, the numerator can be written as
\begin{equation}
\begin{gathered}
\mathcal{N}_{\bar{i}}(\mathbf{k},\mathbf{s})=\sum_{j=1}^n \mathcal{N}_{\bar{i}-\bar{e}^{j}}(\mathbf{k},\mathbf{s})D_j+\Delta_{\bar{i}-\bar{e}^{j}}(\mathbf{k},\mathbf{s}),\quad e^j_k=\delta^j_k.
\end{gathered}
\end{equation}
The polynomial division can be iterated for every $\mathcal{N}_{\bar{i}-\bar{e}^j}(\mathbf{k},\mathbf{s})$. After the complete division of the numerator, the integrand is cast into the following form
\begin{equation}
\frac{\mathcal{N}_{\bar{i}}(\mathbf{k},\mathbf{s})}{D^{i_1}_1\cdots D^{i_n}_n}=\sum_{|j|=0}^{|\overline{i}|}\sum_{\bar{j}}\frac{\Delta_{\bar{j}}(\mathbf{k},\mathbf{s})}{D^{j_1}_1\cdots D^{j_n}_n}.
\end{equation}
The \emph{residues} $\Delta_{\bar{j}}(\mathbf{k},\mathbf{s})$ are polynomials depending of the \emph{irreducible scalar products}, a minimal set of scalar product between loop and external momenta. Therefore, the total double-virtual amplitude can be cast into the form
\begin{equation}
2\text{Re}[\mathcal{M}_{0}^*\mathcal{M}_{2l}]=\sum_{\bar{i}}\sum_{|j|=0}^{|\overline{i}|}\sum_{\bar{j}}A_{\bar{i}}\int_{\mathbf{k}}\frac{\Delta_{\bar{j}}(\mathbf{k},\mathbf{s})}{D^{j_1}_1\cdots D^{j_n}_n}.
\end{equation}

AID has been successfully applied to many amplitude \citep{Mastrolia_2016,mastrolia2016adaptive,torresphdth}. An automated \textsc{Mathematica} implementation of this algorithm has been developed, called \textsc{Aida} \citep{torresphdth,primophdth}. It accepts the \textsc{FeynArts} \citep{HAHN2001418} output (namely the complete unreduced amplitude) and perform the AID, exploiting the \emph{finite fields reconstruction} made available through \textsc{FiniteFlow} \citep{Peraro:2019svx}. Its output is finally translate into integral notation.

\subsection{Integration-by-parts identities}

As a consequence of the $d$-dimensionality and the shift/rotational invariance of the Feynman integrals, an entire class of new relations \citep{Chetyrkin:1981qh} can be found:
\begin{equation}
\begin{gathered}
\int_k f(k)=\int_k e^{v^\mu\frac{\partial}{\partial k^\mu}}f(k)\\
\Downarrow\\
\int_k \frac{\partial}{\partial k^\mu}[v^\mu f(k)]=0,
\end{gathered}
\end{equation}
where $v^\mu$ can be either a loop or an external momenta. These identities between integrals are called \emph{Integration-By-Parts Identities} (IBPs) \cite{Chetyrkin:1981qh,Laporta_2000,GROZIN_2011}. Letting $f(k)$ be the reduced integrands coming from AID,
\begin{equation}
\int_{\mathbf{k}} \frac{\partial}{\partial k_i^\mu}\left[v^\mu \frac{\Delta_{\bar{j}}(\mathbf{k},\mathbf{s})}{D^{j_1}_1\cdots D^{j_n}_n}\right]=0.
\end{equation}
IBPs are employed to generate a large system of equalities, which are not all independent. As a consequence, there exists a minimal set of integrals which can not be reduced further: they are known as \emph{Master Integrals} (MIs) \citep{Smirnov:2010hn}. The general expression for a MI is
\begin{equation}
\mathcal{T}_i(\mathbf{s})=\int_{\mathbf{k}}\frac{S_1^{s_{i1}}\cdots S_m^{s_{im}}}{D^{r_{i1}}_1\cdots D^{r_{in}}_n},
\end{equation}
where $S_1\dots S_m$ are the irreducible scalar products. IBPs allow the ultimate decomposition of the Feynman integrals in the basis of MIs,
\begin{equation}\label{aida}
\int_{\mathbf{k}}\frac{\Delta_{\bar{j}}(\mathbf{k},\mathbf{s})}{D^{j_1}_1\cdots D^{j_n}_n}=\sum_{l=1}^{N_{\text{MIs}}}c_{\bar{j}l}\mathcal{T}_l(\mathbf{s}),
\end{equation}
which amounts the complete double-virtual contribution in this simple form,
\begin{equation}\label{miexp}
\begin{aligned}
2\text{Re}[\mathcal{M}_{0}^*\mathcal{M}_{2l}]&=\sum_{l=1}^{N_{\text{MIs}}}\left[\sum_{\bar{i}}\sum_{|j|=0}^{|\overline{i}|}\sum_{\bar{j}}A_{\bar{i}}c_{\bar{j}l}\right]\mathcal{T}_l(\mathbf{s}).\\
&=\sum_{l=1}^{N_{\text{MIs}}}C_l\mathcal{T}_l(\mathbf{s}).
\end{aligned}
\end{equation}
The choice of the MIs is not unique. Laporta algorithm \citep{Laporta_2000} is an automatable way to choose MIs which respect a "simplicity" criterion. Although Laporta MIs look simpler, they might be not the convenient choice if one wants to evaluate them analytically \citep{Henn_2013,Lee_2015,Argeri:2014qva}. 

There exist several public codes \citep{Smirnov_2008,Georgoudis_2017,Lee_2014} which can provide IBPs and the Laporta MIs for any input topology, such as \textsc{reduze} \citep{manteuffel2012reduze} and \textsc{kira} \citep{Maierhoefer:2017hyi}. It is important to underline that these codes may accept as input a custom MIs basis; in this case, their output will be the IBPs expressed in terms of the chosen basis.

\section{Evaluating the Master Integrals}\label{eval}

The $\epsilon$-expansion of MIs can be achieved by either analytical or numerical evaluation. In this picture, $d=4-2\epsilon$ and the expansion reads
\begin{equation}\label{sol}
\mathcal{T}_i(\mathbf{s})=\sum_{k=-n_p}^\infty\epsilon^k g_k(\mathbf{s}),
\end{equation}
where $n_p$ is the order of the higher order pole into the $\mathcal{T}_i(\mathbf{s})$ series. In practice, the series expansion will be truncated to a finite order, such that every Feynman amplitude is expanded up to $\mathcal{O}(\epsilon)$. Here, the evaluation strategies employed in this calculation are presented.

\subsection{Differential equations}

The analytical approach exploits the IBPs to build a system of differential equation \citep{Barucchi_1973,Kotikov:1990kg,Remiddi:1997ny,Gehrmann_2000} (see also for review \citep{ARGERI_2007,Henn_2015}):
\begin{equation}\label{sys}
\begin{aligned}
\frac{\partial}{\partial s_j}\mathcal{T}_i(s_j)=&[\mathbb{A}(\epsilon,s_j)]_{ik}\mathcal{T}_k(s_j)\\
=&[\mathbb{A}_{0}(s_j)+\epsilon\mathbb{A}_{1}(s_j)]_{ik}\mathcal{T}_k(s_j)
\end{aligned}
\end{equation}
where the MIs basis is chosen in such a way that the matrix $\mathbb{A}(\epsilon,s_j)$ depends linearly on $\epsilon$. 

If there exist a rotation matrix $\mathbb{R}(s_j)$ such that Eq. (\ref{sys}) can be cast into the \emph{canonical form} \citep{Henn_2015}, simbolically expressed as
\begin{gather}
\mathcal{F}_i(s_j)=[\mathbb{R}(s_j)]_{ik}\mathcal{T}_k(s_j)\\
\Downarrow\nonumber\\
\frac{\partial}{\partial s_j}\mathcal{F}_i(s_j)=\epsilon[\mathbb{B}(s_j)]_{ik}\mathcal{F}_k(s_j)
\end{gather}
the PDE admits a solution in terms of the \emph{Chen's iterated integral}
\begin{equation}
\mathcal{F}_i(s_j)=\left(\text{P}e^{\epsilon\int_\gamma d\mathbb{B}_{j}}\right)_{ik}\mathcal{F}_k(s_{j0})
\end{equation}
and its $\epsilon$-expansion casts $\mathcal{F}_i(s_j)$ to the form (\ref{sol}). The boundary condition $\mathcal{F}_k(s_{j0})$ are chosen by exploiting regularity conditions into the phase space appropriate kinematic limits. 

The matrix $\mathbb{R}(s_j)$ can be found by means of the \emph{Magnus exponential method} \citep{Argeri:2014qva,Magnus:1954zz,Blanes_2009}, that provides a close formula for the rotation matrix,
\begin{equation}
\begin{gathered}
\mathbb{R}(s_j)=\exp[\Omega(s_j,s_{j0})],\\
\Omega(s_j,s_{j0})=\sum_{n=1}^\infty \Omega_{n}(s_j,s_{j0}),\\
\begin{aligned}
\Omega_{1}(s_j,s_{j0})&=\int_{s_{j0}}^{s_j}d\tau_1 \mathbb{A}_0(\tau_1),\\
\Omega_{2}(s_j,s_{j0})&=\frac{1}{2}\int_{s_{j0}}^{s_j}d\tau_1\int_{s_{j0}}^{\tau_1}d\tau_2 [\mathbb{A}_0(\tau_1),\mathbb{A}_0(\tau_2)].
\end{aligned}
\end{gathered}
\end{equation}
The complete order-by-order expression for $\Omega(s_j,s_{j0})$ can be found in \citep{Argeri:2014qva}.

The PDE method has been employed in the evaluation of several two-loop Feynman integrals involving a massive fermion-pair \citep{Mastrolia_2017,Di_Vita_2018,Bonciani_2008,Bonciani_2009,Bonciani_2011,Di_Vita_2017,Lee:2019lno,Becchetti:2019tjy,DiVita:2019lpl,Mastrolia:2018sso}. In particular, within our
collaboration, the complete  computation of the MIs for the
$\mu e-$scattering was presented in Refs. \citep{Mastrolia_2017,Di_Vita_2018,Mastrolia:2018sso}.


\subsection{Sector decomposition}

The alternative way to obtain (\ref{sol}) involves numerical methods. 

A general algorithm valid for any multi-loop Feynman integral is the \emph{sector decomposition} \citep{HEINRICH_2008}. Briefly, Feynman parametrization applied $\mathcal{T}_i(\mathbf{s})$ changes the integration space from the Minkowski (or Euclidean) space to the $n$-dimensional unit cube\footnote{$\mathbf{x}=(x_1,\dots,x_n)$} $\mathcal{C}^n$
\begin{equation}
\mathcal{T}_i(\mathbf{s})=\int_{\mathbb{M}} d^dk f_i(k)=\int_{\mathcal{C}^n} d^n\mathbf{x} \tilde{f}_i(\mathbf{x}).
\end{equation}
The unit cube $\mathcal{C}^n$ can be iteratively decomposed and deformed into multiple sub-domains, until the pole structure of the integral manifests multiplicatively into the integrand,
\begin{equation}
\int_{\mathcal{C}^n} d^n\mathbf{x} \tilde{f}_i(\mathbf{x})=\sum_{j=1}^n \int_{\mathcal{C}^n} d^n\mathbf{x} \frac{\hat{f}_i(\mathbf{x})}{x_j^{a_j-b_j\epsilon}}, 
\end{equation}
and subsequently isolated by means of the \emph{end-point subtraction} method. For $a_j=b_j=1$, such method reads
\begin{small}
\begin{equation}
\int_0^1 dx_j \frac{\hat{f}_i(x_j)}{x_j^{1-\epsilon}}=\frac{\hat{f}(0)}{\epsilon}+\int_0^1dx_j \frac{\hat{f}(x_j)-\tilde{f}(0)}{x_j^{1-\epsilon}}.
\end{equation}
\end{small}
Once the sector decomposition has been applied, MIs lie to the following form
\begin{equation}\label{secdec}
\mathcal{T}_i(\mathbf{s})=\sum_{k=-n_p}^\infty\sum_{j=1}^n\epsilon^k \int_{\mathcal{C}^{j_k}} d^{j_k}\mathbf{x} \tilde{f}_{ij_k}(\mathbf{x}).
\end{equation}
The integrals appearing in Eq. (\ref{secdec}) are \emph{finite} and they can be integrated using numerical algorithms. 

The integration on the physical phase-space region usually requires \emph{contour deformation} procedures, such that the threshold singularities of the integrand can be avoided and the numerical stability is preserved.

The complete algorithm has been implemented into a code, \textsc{SecDec} \citep{Borowka_2015}. It goes through the interation of the Sector Decomposition algorithm and performs the numerical evaluation using Monte Carlo integration methods available into the \textsc{Cuba} libraries \citep{Hahn_2005}, with the possibility of deform the integration contour.

\section{Automation}\label{autom}

A complete automation of the evaluating algorithm presented in Sections \ref{dec}. and \ref{eval}. has been developed. Due to the symbolic structure of the amplitude, \textsc{Mathematica} has been chosen to be the main workstation. The specific codes performing the decomposition and the evaluation of the amplitude have been embedded into a \textsc{Mathematica} script.

The generation of the double-virtual contribution is carried out by \textsc{FeynArts} and \textsc{FeynCalc}, which provide the raw input for the decomposition algorithms. In particular, \textsc{FeynCalc} performs the Dirac and tensor algebra and deals with the eventual Dirac traces.

The amplitude is now ready to be decomposed at integrand level by means of \textsc{Aida}, obtaining the structure expressed in (\ref{aida}). The new integrands are then converted into integrals, by a convenient notation.

Integrals belonging to the amplitude are given as input to \textsc{Reduze} or \textsc{Kira}, which perform the IBP reduction. An interface has been build, that automates the configuration of \textsc{Reduze} and reads its output into a \textsc{Mathematica} database file. At this stage, the double-virtual amplitude takes the form (\ref{miexp}).

The last step is the evaluation of the MIs. Analytical values of the MIs is stored into \textsc{Mathematica} package \citep{Mastrolia_2017,Di_Vita_2018}. It automatically replaces the symbolic integrals to its $\epsilon$-pole Laurent expansion. 

Alternatively, an additional interface with \textsc{SecDec} has been developed. It automatically provides its input files and automate the numerical evaluation. The output of \textsc{SecDec} is then stored into a \textsc{Mathematica} package. To check the consistency of the result, both of the approached have been used. 

\begin{figure}
\centering 
\includegraphics[scale=.3]{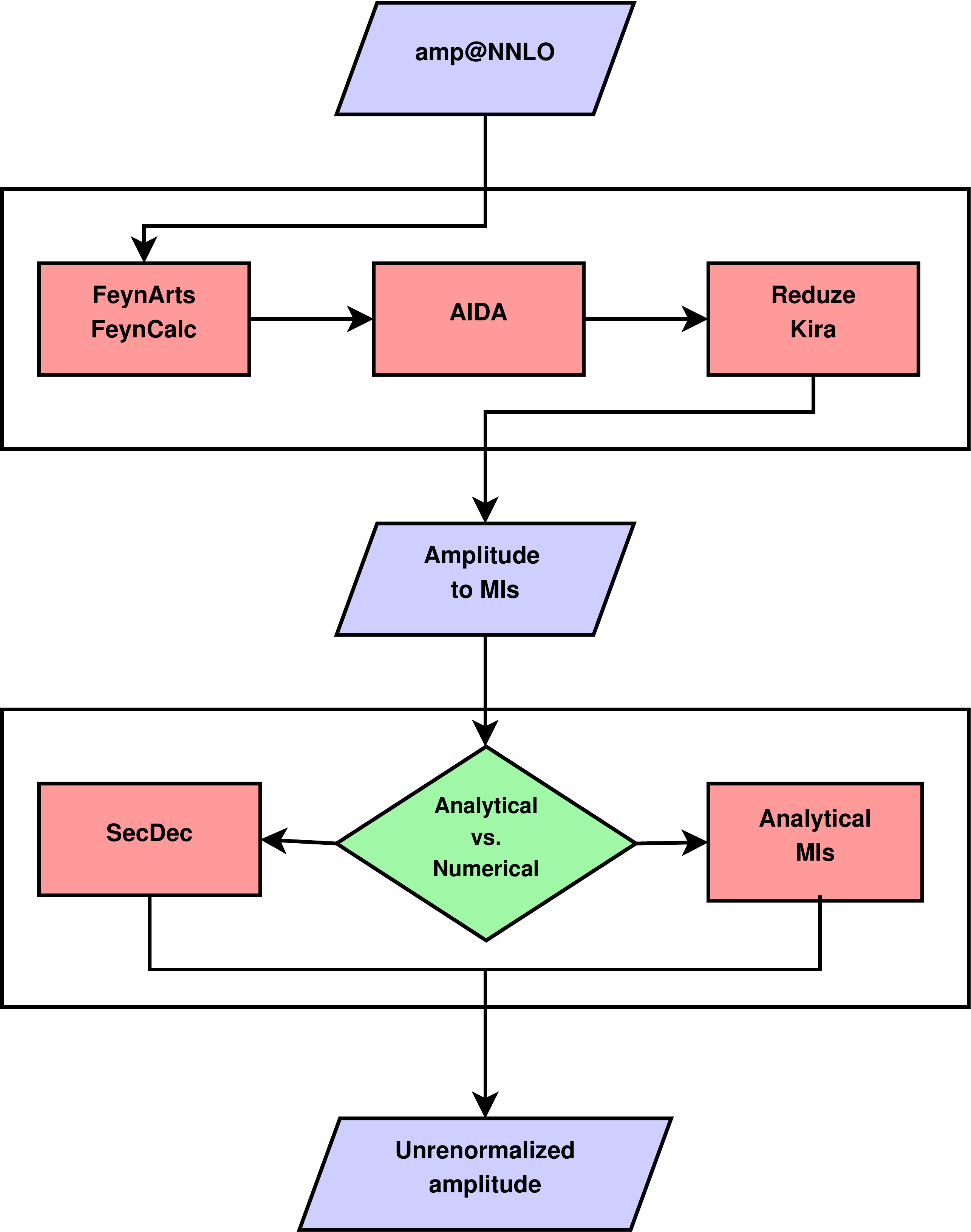}
\caption{Flowchart of the algorithm}\label{fig}
\end{figure}

The algorithm here presented is actually completely general, and this calculation represents a strong check of the validity of the method. A flowchart representing the data flow is given in Fig. \ref{fig}.

\section{Renormalization}\label{renorm}

QED is a renormalizable theory. As a consequence, UV divergencies arising from divergent loop integrals can be regularized and canceled order-by-order in the coupling expansion. 

A diagrammatic approach to the renormalization is the most convenient way to obtain a UV finite amplitude out of the NNLO scattering amplitude; renormalized perturbation theory will be the ideal framework.

In order to renormalize QED, we consider the QED Lagrangian with $N_f=1+1$ active flavor, one massive (muon) and one massless (electron) \citep{Bohm:2001yx}:
\begin{equation}
\begin{aligned}
\mathcal{L}_0&=-\frac{1}{4}F^{\mu\nu}_0 F_{0,\mu\nu}+\frac{1}{2\xi_0}(\partial_\mu A_0^\mu)^2\\
&+\overline{\psi}_{0}(i\slashed{p}-m_0)\psi_{0}+\overline{\chi}_{0}(i\slashed{p})\chi_{0}\\
&+e_0A^\mu_0(\overline{\psi}_{0}\gamma_\mu\psi_{0}+\overline{\chi}_{0}\gamma_\mu\chi_{0}).
\end{aligned}
\end{equation}
In this approach, fields, couplings and masses are treated to be \emph{bare} (denoted with the subscript 0). Bare quantities are connected to the \emph{renormalized} ones by
\begin{equation}
\begin{gathered}
\psi_{0}=Z_\psi^{\frac{1}{2}} \psi, \quad \chi_{0}=Z_\chi^{\frac{1}{2}} \chi,\\
A^\mu_0=Z_A^{\frac{1}{2}} A^\mu, \quad \xi_0=Z_\xi \xi,\\
e_0=Z_e e, \quad m_0=Z_m m.
\end{gathered}
\end{equation}
The bare Lagrangian can be expressed in terms of the renormalized quantites:
\begin{equation}
\mathcal{L}_0=\mathcal{L}_R+\mathcal{L}_{\text{ct}},
\end{equation}
where
\begin{equation}
\begin{aligned}
\mathcal{L}_R=&-\frac{1}{4}F^{\mu\nu} F_{\mu\nu}+\frac{1}{2\xi}(\partial_\mu A^\mu)^2\\
&+\overline{\psi}(i\slashed{p}-m)\psi+\overline{\chi}(i\slashed{p})\chi\\
&+eA^\mu(\overline{\psi}\gamma_\mu\psi+\overline{\chi}\gamma_\mu\chi),
\end{aligned}
\end{equation}
and
\begin{equation}
\begin{aligned}
\mathcal{L}_{\text{ct}}=&-\frac{1}{4}(Z_A-1)F^{\mu\nu} F_{\mu\nu}\\
&+\frac{1}{2\xi}(Z_AZ_\xi^{-1}-1)(\partial_\mu A^\mu)^2 \\
&+(Z_\psi-1)\overline{\psi}(i\slashed{p})\psi+(Z_\chi-1)\overline{\chi}(i\slashed{p})\chi\\
&-(Z_\psi Z_m-1) m\overline{\psi}\psi\\
&+(Z_\psi Z_A^{\frac{1}{2}} Z_e-1)eA^\mu(\overline{\psi}\gamma_\mu\psi)\\
&+(Z_\chi Z_A^{\frac{1}{2}} Z_e-1)eA^\mu(\overline{\chi}\gamma_\mu\chi).
\end{aligned}
\end{equation}
The \emph{counterterm Lagrangian} $\mathcal{L}_{\text{ct}}$ contains the divergent behaviour arising from the bare quantities. 

Ward Identities in QED constrain ratios between the normalizations $Z_i$ \citep{Bohm:2001yx}. 
\begin{gather}\label{wi}
\frac{Z_A}{Z_\xi}=\kappa_\xi,\quad 
\frac{Z_A^{\frac{1}{2}}}{Z_\xi Z_e}=\kappa_e,
\end{gather}
where $\kappa_\xi$ and $\kappa_e$ are arbitrary \emph{finite} constants. By choosing $\kappa_i=1$, the normalizations become
\begin{equation}\label{wi2}
Z_\xi=Z_A,\quad 
Z_e=Z_A^{-\frac{1}{2}}.
\end{equation}
Note that the choice of the renormalization scheme for the photon field $A^\mu$ leads automatically to fix the same prescription to the coupling $e$. This is a direct consequence of the Ward Identities (\ref{wi2}) in QED .

It is important to stress that Eq. (\ref{wi}) constrains only the divergent part of the normalizations, letting their finite part unconstrained. Hence,  two different renormalization prescription for $Z_e$ and $Z_A$ can be chosen, e.g. $Z_A$ in \emph{on-shell} scheme and $Z_e$ in $\overline{\text{MS}}$ scheme. This alternative choice brings the Ward identity to be
\begin{equation}
Z_e=Z_A^{-\frac{1}{2}}\bigr|_{\text{finite}=0} ,
\end{equation}
which yields no contradiction with Eq. (\ref{wi}).

Imposing the Ward identities (\ref{wi}), the counterterms Lagrangian becomes
\begin{equation}
\begin{aligned}
\mathcal{L}_{\text{ct}}=&-\frac{1}{4}(Z_A-1)F^{\mu\nu} F_{\mu\nu}\\
&+(Z_\psi-1)\overline{\psi}(i\slashed{p})\psi+(Z_\chi-1)\overline{\chi}(i\slashed{p})\chi\\
&-(Z_\psi Z_m-1) m\overline{\psi}\psi\\
&+(Z_\psi -1)eA^\mu(\overline{\psi}\gamma_\mu\psi)\\
&+(Z_\chi -1)eA^\mu(\overline{\chi}\gamma_\mu\chi).
\end{aligned}
\end{equation}
Finally, by setting $Z_i=1+\delta _i$
\begin{equation}
\begin{aligned}
\mathcal{L}_{\text{ct}}=&-\frac{1}{4}\delta_AF^{\mu\nu} F_{\mu\nu} \\
&+\delta_\psi\overline{\psi}(i\slashed{p})\psi+\delta_\chi\overline{\chi}(i\slashed{p})\chi\\
&-(\delta_\psi+\delta_m+\delta_\psi\delta_m) m\overline{\psi}\psi\\
&+\delta_\psi eA^\mu(\overline{\psi}\gamma_\mu\psi)+\delta_\chi eA^\mu(\overline{\chi}\gamma_\mu\chi).\\
\end{aligned}
\end{equation}
The Ward identities have been chosen in such a way that only field and mass renormalization is needed. The coupling constant $e$ is automatically renormalized thanks to the Ward identities. 

The counterterm Lagrangian provides additional Feynman rules, which lead to UV finite amplitude. Physical results can only be obtained by choosing a \emph{renormalization scheme} that fixes the residual unphysical degrees of freedom. 

Different renormalization schemes bring changes to the 4-point Green function. 
The $S-$matrix element $\bra{e\mu}S\ket{e\mu}$ is related to the \emph{amputated renormalized} Green function \mbox{$\bra{0}T\{\psi(p_1)\chi(p_2)\overline{\psi}(p_3)\overline{\chi}(p_4)\}\ket{0}_{amp}=G_4$} by
\begin{equation}
\begin{aligned}
\bra{\mu e}S\ket{\mu e}&=(\slashed{p}_1-m_{P})\slashed{p}_2\slashed{p}_3(\slashed{p}_4-m_{P})G_4,
\end{aligned}
\end{equation}
known as Lehmann-Symanzik-Zimmermann (LSZ) reduction formula \citep{Lehmann1955,Schwartz:2013pla}. The Dirac operators \mbox{$(\slashed{p}_i-m_{P})$} contain the wave-function corrections, and $m_{P}$ is  the \emph{pole mass} of the muon. Different renormalization scheme choices lead to different renormalized masses $m$, introducing a \emph{residue} at the pole mass $R_\psi\neq 1$,

\begin{equation}
\begin{gathered}
\lim_{\slashed{p}\to m_P}(\slashed{p}-m_P)\frac{i}{\slashed{p}-m}=R_\psi\\
\Downarrow\\
\bra{0}T\{\psi(p)\psi(-p)\}\ket{0}=\frac{i R_\psi}{\slashed{p}-m_P}+\text{regular terms}.
\end{gathered}
\end{equation}
The LSZ reduction formula changes according to the new residue, acquiring new factors \citep{Bohm:2001yx}:
\begin{equation}
\begin{aligned}
\bra{\mu e}S\ket{\mu e}&=(\slashed{p}_1-m_{P})\slashed{p}_2\slashed{p}_3(\slashed{p}_4-m_{P})\frac{G_4}{R_\psi R_\chi}.
\end{aligned}
\end{equation}
Expressing the residue as $R_i=1+\delta R_i$, 
the order-by-order structure of the renormalized Green function become manifest
\begin{equation}
\begin{gathered}
G_4=G_{\text{tree}}+G_{1l}+G_{2l}+\cdot,\\
\delta R_i=\delta R_{i,1l}+\delta R_{i,2l}+\cdots,
\end{gathered}
\end{equation}
which yields to the following structure
\begin{equation}
\begin{aligned}
G_4&=G_{\text{tree}}\\
&+[G_{1l}-(\delta R_{\psi,1l}+\delta R_{\chi,1l})G_{\text{tree}}]\\
&+[G_{2l}-(\delta R_{\psi,1l}+\delta R_{\chi,1l})G_{1l}\\
&-(\delta R_{\psi,2l}+\delta R_{\chi,2l})G_{\text{tree}}+\delta R_{\psi,1l}\delta R_{\chi,1l}G_{\text{tree}}].
\end{aligned}
\end{equation}
The choice of the \emph{on-shell} renormalization scheme for the muon mass and fields sets \mbox{$\delta R_{\psi,i}=0$} identically.

Therefore, the renormalization scheme choice applied in the $\mu e\to \mu e$ scattering is:
\begin{itemize}
\item[-] $\overline{\text{MS}}$ scheme for the \emph{coupling}, the \emph{photon} and the \emph{electron} fields
\item[-] On-shell scheme for the \emph{muon} field and mass
\end{itemize} 
The muon field is treated differently due to its non-vanishing mass. Despite of the complications it introduces at integration level, the opportunity to use on-shell scheme for the muon yields simplification at renormalization level.

\begin{figure*}
\centering
\begin{minipage}{.45\textwidth}
\includegraphics[width=\textwidth,height=9cm]{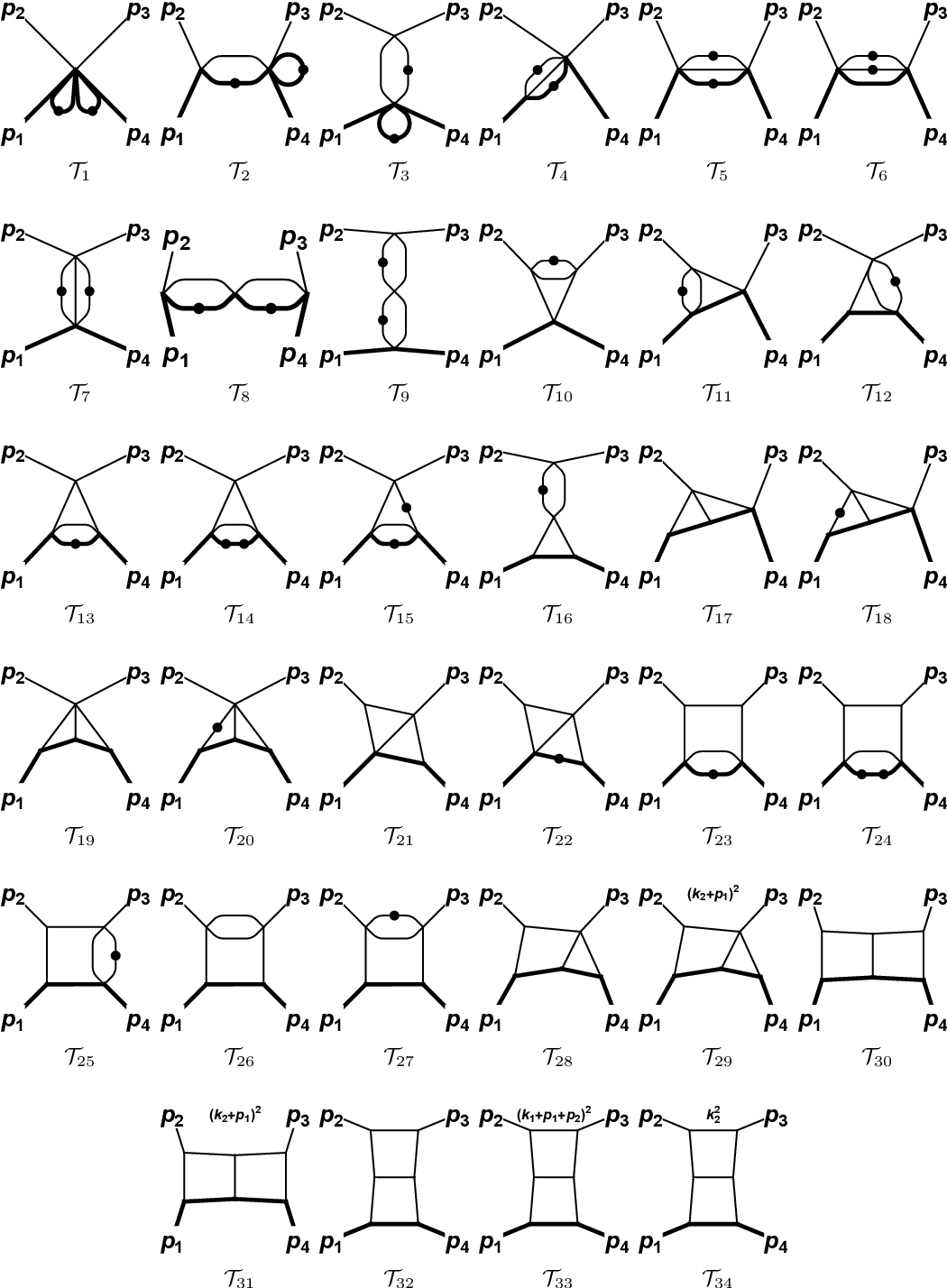}
\end{minipage}
\begin{minipage}{.05\textwidth}
~
\end{minipage}
\begin{minipage}{.45\textwidth}
\includegraphics[width=\textwidth,height=10cm]{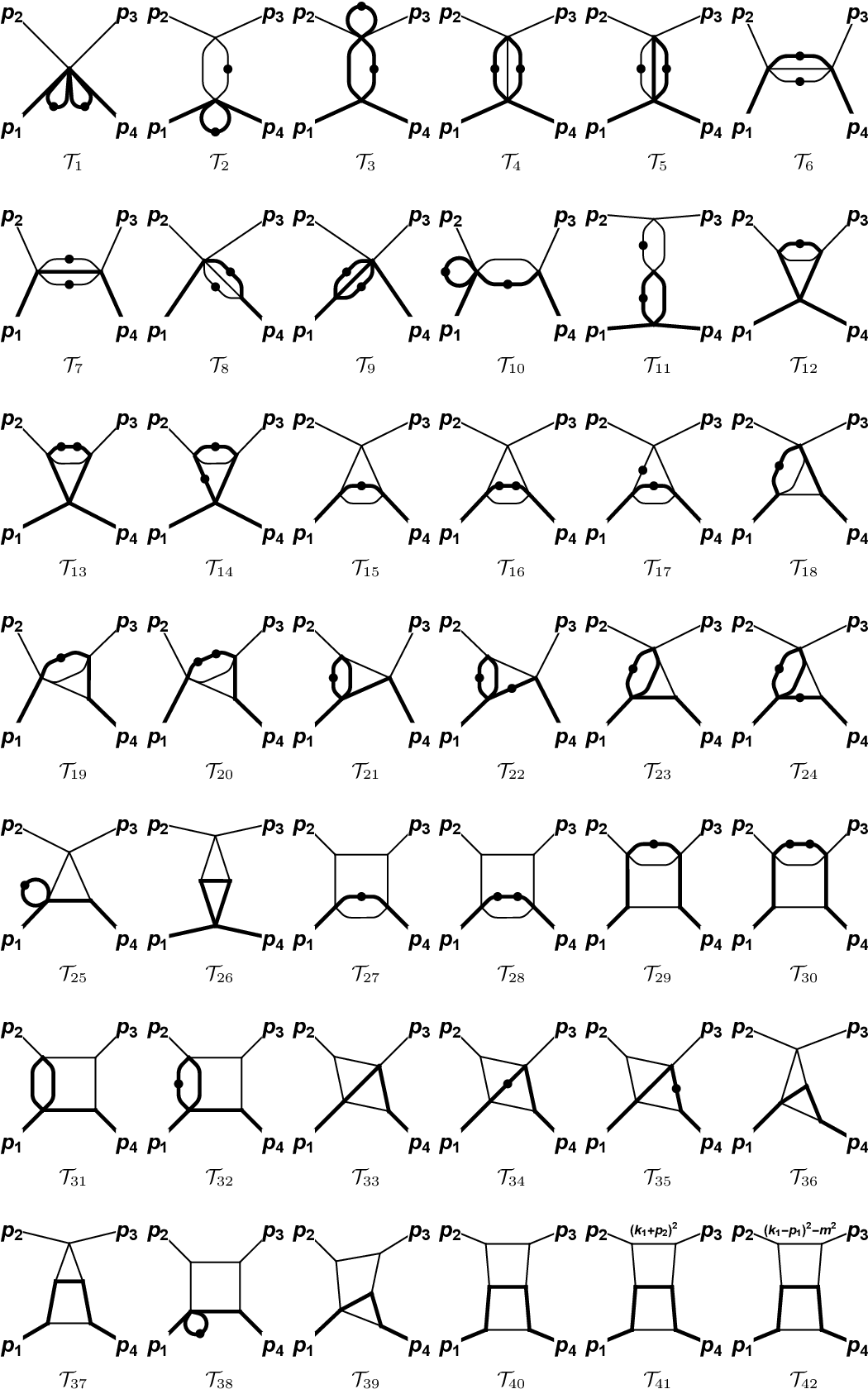}
\end{minipage}

\par\bigskip
\par\bigskip
\begin{minipage}{.45\textwidth}
\includegraphics[width=\textwidth,height=12cm]{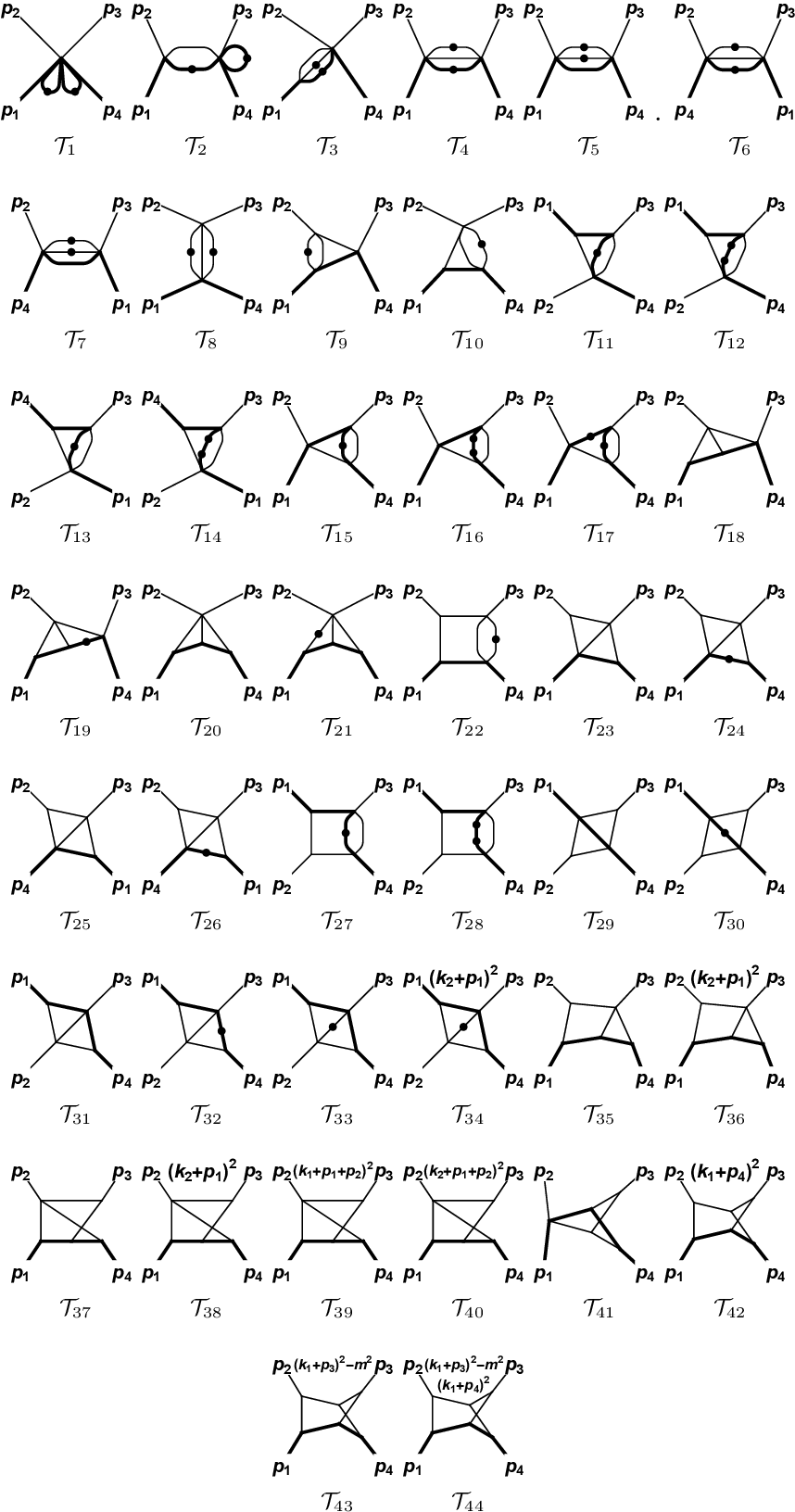}
\end{minipage}

\par\bigskip
\caption{Master Integrals belonging to planar and non-planar topologies for the QED NNLO $\mu e$-scattering amplitude.}\label{mis}
\end{figure*}

\section{Summary}
There are 69 two-loop Feynman diagrams contributing to the QED NNLO $\mu e$-scattering process (Fig. \ref{diags}). The complete amplitude gets contributions from $\sim 10^4$ Feynman integrals with maximum rank equal to 4. 

AID and IBPs reduction with respectively \textsc{Aida} and \textsc{Kira} have decreased significantly the number of integrals \citep{Mastrolia_2017,Di_Vita_2018}, providing $N_{MI}=120$ with max rank equal to 2 (Fig. \ref{mis}). These MIs are the irreducible set of integrals that one has to evaluate and further reduction of the rank is not possible. The presence of irreducible scalar products, typical of two-loop topology, does not allow a reduction to scalar (rank-0) integrals. 

The differential equation method has allowed the analytical evaluation of the planar \citep{Mastrolia_2017} and non-planar \citep{Di_Vita_2018} sets of MIs. The consistency of the result has been checked by evaluating the MIs with \textsc{GiNaC} and comparing the result with the numerical integration with \textsc{SecDec}, finding agreement between the two approaches.

At the present state of the calculation, the analytical \emph{un}renormalized amplitude has been achieved. The renormalization strategy discussed in Section \ref{renorm} is still in progress. A careful check of the  cancellation of the UV divergencies is required in order to obtain a UV finite contribution. 
\par\bigskip
The double-virtual contribution represents the most challenging part of the full $\mu e$-scattering cross section. However, the complete UV/IR finite NNLO amplitude can be achieved only by including all the contributions in Eq. (\ref{completecalc}). A particular care should be given to the soft and collinear limit due to photon emission from the electron, which in our approximation are treated as massless. 

Real-virtual and double-real correction involve 1-loop and tree-level calculations, which can be dealt with the same technology shown in Section \ref{autom}. For 1-loop amplitudes, additional tools are available on the market e.g. \textsc{GoSam} \cite{Cullen2012} or \textsc{Package-X} \cite{Patel:2015tea}.

\section{Acknowledgment}

This work has been supported by the Generalitat Valenciana, the Spanish Government and ERDF funds from the EU Commission and the Spanish Centro de Excelencia Severo Ochoa Programme (Grants No. SEJI/2017/019, FPA2017-84445-P and SEV-2014-0398). 

%
%

\bibliography{mybib}

\begin{thebibliography}{58}

\bibitem{blum2013muon}
T.~Blum, A.~Denig, I.~Logashenko, E.~de~Rafael, B.L. Roberts, T.~Teubner,
  G.~Venanzoni, \emph{The muon (g-2) theory value: Present and future} (2013),
  \texttt{1311.2198}

\bibitem{Carloni_Calame_2015}
C.~Carloni~Calame, M.~Passera, L.~Trentadue, G.~Venanzoni, Physics Letters B
  \textbf{746}, 325–329 (2015)

\bibitem{Abbiendi_2017}
G.~Abbiendi, C.M.C. Calame, U.~Marconi, C.~Matteuzzi, G.~Montagna,
  O.~Nicrosini, M.~Passera, F.~Piccinini, R.~Tenchini, L.~Trentadue et~al., The
  European Physical Journal C \textbf{77} (2017)

\bibitem{Alacevich:2018vez}
M.~Alacevich, C.M. Carloni~Calame, M.~Chiesa, G.~Montagna, O.~Nicrosini,
  F.~Piccinini, JHEP \textbf{02}, 155 (2019), \texttt{1811.06743}

\bibitem{Fael:2018dmz}
M.~Fael, JHEP \textbf{02}, 027 (2019), \texttt{1808.08233}

\bibitem{Fael:2019nsf}
M.~Fael, M.~Passera, Phys. Rev. Lett. \textbf{122}, 192001 (2019),
  \texttt{1901.03106}

\bibitem{Mastrolia_2016}
P.~Mastrolia, T.~Peraro, A.~Primo, Journal of High Energy Physics \textbf{2016}
  (2016)

\bibitem{mastrolia2016adaptive}
P.~Mastrolia, T.~Peraro, A.~Primo, W.J. Torres~Bobadilla, PoS \textbf{LL2016},
  007 (2016), \texttt{1607.05156}

\bibitem{primophdth}
A.~Primo, Ph.D. thesis, Padua U. (2017)

\bibitem{torresphdth}
W.J.T. Bobadilla, Ph.D. thesis, Padua U. (2017)

\bibitem{Chetyrkin:1981qh}
K.G. Chetyrkin, F.V. Tkachov, Nucl. Phys. \textbf{B192}, 159 (1981)

\bibitem{Laporta_2000}
Laporta, International Journal of Modern Physics A \textbf{15}, 5087 (2000)

\bibitem{manteuffel2012reduze}
A.~von Manteuffel, C.~Studerus, \emph{Reduze 2 - distributed feynman integral
  reduction} (2012), \texttt{1201.4330}

\bibitem{Maierhoefer:2017hyi}
P.~Maierhöfer, J.~Usovitsch, P.~Uwer, Comput. Phys. Commun. \textbf{230}, 99
  (2018), \texttt{1705.05610}

\bibitem{Barucchi_1973}
G.~Barucchi, G.~Ponzano, Journal of Mathematical Physics \textbf{14}, 396
  (1973)

\bibitem{Kotikov:1990kg}
A.V. Kotikov, Phys. Lett. \textbf{B254}, 158 (1991)

\bibitem{Remiddi:1997ny}
E.~Remiddi, Nuovo Cim. \textbf{A110}, 1435 (1997), \texttt{hep-th/9711188}

\bibitem{Gehrmann_2000}
T.~Gehrmann, E.~Remiddi, Nuclear Physics B \textbf{580}, 485–518 (2000)

\bibitem{Mastrolia_2017}
P.~Mastrolia, M.~Passera, A.~Primo, U.~Schubert, Journal of High Energy Physics
  \textbf{2017} (2017)

\bibitem{Di_Vita_2018}
S.~Di~Vita, S.~Laporta, P.~Mastrolia, A.~Primo, U.~Schubert, Journal of High
  Energy Physics \textbf{2018} (2018)

\bibitem{HEINRICH_2008}
G.~Heinrich, International Journal of Modern Physics A \textbf{23}, 1457–1486
  (2008)

\bibitem{Ossola_2007}
G.~Ossola, C.G. Papadopoulos, R.~Pittau, Nuclear Physics B \textbf{763},
  147–169 (2007)

\bibitem{Mastrolia_2012}
P.~Mastrolia, E.~Mirabella, T.~Peraro, Journal of High Energy Physics
  \textbf{2012} (2012)

\bibitem{Badger_2012}
S.~Badger, H.~Frellesvig, Y.~Zhang, Journal of High Energy Physics
  \textbf{2012} (2012)

\bibitem{TKACHOV198165}
F.~Tkachov, Physics Letters B \textbf{100}, 65  (1981)

\bibitem{Passarino:1978jh}
G.~Passarino, M.J.G. Veltman, Nucl. Phys. \textbf{B160}, 151 (1979)

\bibitem{Smirnov_2006}
A.V. Smirnov, V.A. Smirnov, Journal of High Energy Physics \textbf{2006},
  001–001 (2006)

\bibitem{vanNeerven:1983vr}
W.L. van Neerven, J.A.M. Vermaseren, Phys. Lett. \textbf{137B}, 241 (1984)

\bibitem{Mastrolia_20122}
P.~Mastrolia, E.~Mirabella, G.~Ossola, T.~Peraro, Physics Letters B
  \textbf{718}, 173–177 (2012)

\bibitem{HAHN2001418}
T.~Hahn, Computer Physics Communications \textbf{140}, 418  (2001)

\bibitem{Peraro:2019svx}
T.~Peraro, JHEP \textbf{07}, 031 (2019), \texttt{1905.08019}

\bibitem{GROZIN_2011}
A.G. Grozin, International Journal of Modern Physics A \textbf{26}, 2807–2854
  (2011)

\bibitem{Smirnov:2010hn}
A.V. Smirnov, A.V. Petukhov, Lett. Math. Phys. \textbf{97}, 37 (2011),
  \texttt{1004.4199}

\bibitem{Henn_2013}
J.M. Henn, Physical Review Letters \textbf{110} (2013)

\bibitem{Lee_2015}
R.N. Lee, Journal of High Energy Physics \textbf{2015} (2015)

\bibitem{Argeri:2014qva}
M.~Argeri, S.~Di~Vita, P.~Mastrolia, E.~Mirabella, J.~Schlenk, U.~Schubert,
  L.~Tancredi, JHEP \textbf{03}, 082 (2014), \texttt{1401.2979}

\bibitem{Smirnov_2008}
A.~Smirnov, Journal of High Energy Physics \textbf{2008}, 107–107 (2008)

\bibitem{Georgoudis_2017}
A.~Georgoudis, K.J. Larsen, Y.~Zhang, Computer Physics Communications
  \textbf{221}, 203–215 (2017)

\bibitem{Lee_2014}
R.N. Lee, Journal of Physics: Conference Series \textbf{523}, 012059 (2014)

\bibitem{ARGERI_2007}
M.~Argeri, P.~Mastrolia, International Journal of Modern Physics A \textbf{22},
  4375–4436 (2007)

\bibitem{Henn_2015}
J.M. Henn, Journal of Physics A: Mathematical and Theoretical \textbf{48},
  153001 (2015)

\bibitem{Magnus:1954zz}
W.~Magnus, Commun. Pure Appl. Math. \textbf{7}, 649 (1954)

\bibitem{Blanes_2009}
S.~Blanes, F.~Casas, J.~Oteo, J.~Ros, Physics Reports \textbf{470}, 151–238
  (2009)

\bibitem{Bonciani_2008}
R.~Bonciani, A.~Ferroglia, T.~Gehrmann, D.~Maître, C.~Studerus, Journal of
  High Energy Physics \textbf{2008}, 129–129 (2008)

\bibitem{Bonciani_2009}
R.~Bonciani, A.~Ferroglia, T.~Gehrmann, C.~Studerus, Journal of High Energy
  Physics \textbf{2009}, 067–067 (2009)

\bibitem{Bonciani_2011}
R.~Bonciani, A.~Ferroglia, T.~Gehrmann, A.~von Manteuffel, C.~Studerus, Journal
  of High Energy Physics \textbf{2011} (2011)

\bibitem{Di_Vita_2017}
S.~Di~Vita, P.~Mastrolia, A.~Primo, U.~Schubert, Journal of High Energy Physics
  \textbf{2017} (2017)

\bibitem{Lee:2019lno}
R.N. Lee, K.T. Mingulov (2019), \texttt{1901.04441}

\bibitem{Becchetti:2019tjy}
M.~Becchetti, R.~Bonciani, V.~Casconi, A.~Ferroglia, S.~Lavacca, A.~von
  Manteuffel, JHEP \textbf{08}, 071 (2019), \texttt{1904.10834}

\bibitem{DiVita:2019lpl}
S.~Di~Vita, T.~Gehrmann, S.~Laporta, P.~Mastrolia, A.~Primo, U.~Schubert, JHEP
  \textbf{06}, 117 (2019), \texttt{1904.10964}

\bibitem{Mastrolia:2018sso}
P.~Mastrolia, M.~Passera, A.~Primo, U.~Schubert, W.J. Torres~Bobadilla, EPJ Web
  Conf. \textbf{179}, 01014 (2018)

\bibitem{Borowka_2015}
S.~Borowka, G.~Heinrich, S.~Jones, M.~Kerner, J.~Schlenk, T.~Zirke, Computer
  Physics Communications \textbf{196}, 470–491 (2015)

\bibitem{Hahn_2005}
T.~Hahn, Computer Physics Communications \textbf{168}, 78–95 (2005)

\bibitem{Bohm:2001yx}
M.~Bohm, A.~Denner, H.~Joos, \emph{{Gauge theories of the strong and
  electroweak interaction}} (2001), ISBN 9783519230458, 9783322801623,
  9783322801609

\bibitem{Lehmann1955}
H.~Lehmann, K.~Symanzik, W.~Zimmermann, Il Nuovo Cimento (1955-1965)
  \textbf{1}, 205 (1955)

\bibitem{Schwartz:2013pla}
M.D. Schwartz, \emph{{Quantum Field Theory and the Standard Model}} (Cambridge
  University Press, 2014), ISBN 1107034736, 9781107034730

\bibitem{Cullen2012}
G.~Cullen, N.~Greiner, G.~Heinrich, G.~Luisoni, P.~Mastrolia, G.~Ossola,
  T.~Reiter, F.~Tramontano, The European Physical Journal C \textbf{72}, 1889
  (2012)

\bibitem{Patel:2015tea}
H.H. Patel, Comput. Phys. Commun. \textbf{197}, 276 (2015), \texttt{1503.01469}

\end{thebibliography}

%
%
%
%

\end{document}